%% file: Noise2_xyz.tex
\documentstyle[epsfig,nolta]{article} 

\newcommand{\Ss}{{\bf s}}
\begin{document} 
\title{NONLINEAR PROJECTIVE FILTERING II:\\
APPLICATION TO REAL TIME SERIES}
\author{Thomas Schreiber$^{\dag}$ and Holger Kantz$^{\ddag}$}
\affiliation{\dag Physics Department, University of Wuppertal, 
   D--42097 Wuppertal\\
   {\tt schreibe@theorie.physik.uni-wuppertal.de}\\
   \ddag Max Planck Institute for Physics of Complex Systems,
      N\"othnitzer Str. 38, D--01187 Dresden\\
   {\tt kantz@mpipks-dresden.mpg.de}}

\maketitle
\abstract
   We discuss applications of nonlinear filtering of time series by locally
   linear phase space projections. Noise can be reduced whenever the error due
   to the manifold approximation is smaller than the noise in the
   system. Examples include the real time extraction of the fetal
   electrocardiogram from abdominal recordings.
\endabstract

\section{INTRODUCTION}
For nonlinear signals with intrinsic instabilities, the tasks of noise
reduction --- or signal separation in general --- is difficult to accomplish
with the traditional spectral approach since the signals may exhibit braod band
frequency spectra.  The theory of nonlinear dynamical systems, or chaos theory,
provides alternative methods for these purposes based on a phase space
representation of the data. The theory behind these methods is outlined in
Ref.~\cite{I} in this volume.  For the derivation of time series methods within
the framework of chaos theory, heavy use has been made of the theoretical
properties of nonlinear deterministic systems.  For most methods, this also
severely restricts the scope of systems they can be applied to. Formally, this
is also the case for nonlinear filtering procedures that exploit the peculiar
structure generated by deterministic phase space dynamics.  In this paper we
will show that if these algorithms are used with care, they can also give
superior results in situations when pure determinism cannot be assumed. The
reason is that, when represented in a low dimensional phase space, also
non-deterministic systems may exhibit structures suitable for filtering
purposes. We will give some examples of the latter statement and also discuss
practical issues that have so far hampered widespread use of nonlinear filters.

\section{METHOD}
A scalar time series $\{s_n\}, n=1,\ldots,N$ can be unfolded in a
multi-dimensional effective phase space using time delay coordinates
${\Ss}_n=(s_{n-(m-1)\tau},\ldots,s_n)$ ($\tau$ is a delay time). If $\{s_n\}$
is a scalar observation of a deterministic dynamical system, it can be shown
under certain genericity conditions~\cite{takens,embed} that the reconstructed
point set is a one-to-one image of the original attractor of the dynamical
system.  We will not assume here that there is such an underlying deterministic
system. Nevertheless, general serial dependencies among the $\{s_n\}$ will
cause the delay vectors $\{\Ss_n\}$ to fill the available $m$-dimensional
space in an inhomogeneous way. Linearly correlated Gaussian random variates
will for example be distributed according to an anisotropic multivariate
Gaussian distribution. Linear geometric filtering~\cite{svd} in phase
space seeks to identify the principal directions of this distribution and
project onto them. The present algorithm can be seen as a nonlinear
generalisation of this approach that takes into account that nonlinear signals
will form {\em curved} structures in delay space. In particular, noisy
deterministic signals form smeared-out lower dimensional manifolds. Nonlinear
phase space filtering seeks to identify such structures and project unto them
in order to reduce noise.

Let us recall the three main steps involved in the noise reduction algorithm
described in Ref.~\cite{I}: (1)~Find a low dimensional approximation to the
``attractor'' described by the trajectory $\{\Ss_n \}$. (2) Project each point
$\Ss_n$ in the trajectory orthogonally onto the approximation to the attractor
to produce a cleaned vector $\hat\Ss_n$ (3) Convert the sequence of cleaned
vectors $\hat\Ss_n$ back into the scalar time domain to produce a cleaned time
series $\hat{s}_n$.

\section{REAL TIME FILTERING}
All of the methods that have been proposed in the literature are formulated as
{\sl a posteriori} filters. The whole signal has to be available before a
cleaned version can be computed, which is then invariably quite computer time
intensive. One class of methods uses a global nonlinear function to represent
the dynamics (at least approximately). This function has to be determined by a
delicate fitting procedure and the actual filtering scheme (for example
Ref.~\cite{davies}) consists of an iterative minimisation procedure. The other
class of algorithms approximates the dynamics in phase space, or phase space
geometry, by locally linear mappings. Here the tessellation of phase space into
small neighbourhoods is the most time consuming step, along with the need to
solve a least squares problem in each of these neighbourhoods. With fairly low
dimensional signals, fast neighbour search algorithms (see~\cite{neigh} for an
overview) are very helpful in this regard.  Let us introduce some modifications
to the locally projective noise reduction scheme discussed in Ref.~\cite{I}
that make its use in real time signal processing feasible.  (1) The data base
of local neighbours that is needed to approximate the dynamics is restricted to
points in the past.  As a side effect, the curvature correction can be carried
out during the first sweep through the data. (2) The number of neighbours
required for each point is limited to a number that is just sufficient for
statistical stability. (3) The last modification uses the fact that the
dynamics is supposed to vary smoothly in phase space. The full linearised
dynamics is reduced to a collection of representative points which is stored
together with their local linear structure. Consequently, the local linear
problem has to be solved only for points that are not yet well approximated by
such a representative.

Thus, we have turned the procedure into a causal filter by restricting
neighbour search to points defined by measurements made in the the past,
$k<n$. By further limiting the number of neighbours searched for to $U_{\rm
max}$, we have sped up the formation of the local covariance matrices
considerably. Still, as the algorithm stands, we have to solve an $(m\times m)$
eigenvalue problem for each point that is to be processed. A large fraction of
this work can be avoided on the base of the assumption that the local linear
structure changes smoothly over phase space. By making local linear
approximations we have already assumed smoothness of the underlying manifold in
the $C_1$ sense. In most physical systems, the additional assumption of $C_2$
smoothness is not less justified. We cannot, however expect that the vectors
which span the principal directions vary slowly from point to point. The reason
is that often some eigenvalues are nearly degenerate and change indices from
point to point when they are ordered by their magnitude.  We thus refrain from
interpolating principal components between phase space points.  Instead, we
choose a length scale $h$ in phase space which is small enough such that the
linear subspaces spanned by the local principal components can be regarded as
effectively the same.  Now we successively build up a data base of
representative points for which the local points of tangency
$\overline\Ss^{(n')}$, and the local principal directions ${\bf c}^q,\quad
q=1,\ldots,Q$ have been determined already. For each new point $\Ss_n$ that is
to be processed, we go through this collection of points to determine whether a
representative is available closer than $h$. In that case, we use the stored
tangent point and principal directions of the representative in order to
perform the projections. If not, a neighbourhood is formed around $\Ss_n$ in
which the eigenvalue problem is solved. The point $\Ss_n$ is then included in
the list of representatives.

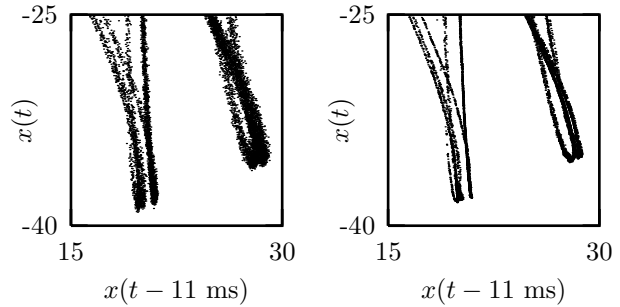
\begin{figure}
\centerline{\hspace*{1.5cm}\input{f_raser.tex}\hspace*{-3cm}\input{f_raser_c.tex}}
\caption[]{\label{fig:raser}Result of nonlinear filtering of an NMR laser time
  series. An enlargement of about one quarter of the total linear extent of
  the attractor is shown.  }
\end{figure}

For real time application, this means that the corrected value $\hat s_n$
cannot be available before $s_{n+m-1}$ has been measured and processed.
Usually, however, this delay window is a very short time, at least compared to
the duration of the recording, and the procedure can be regarded as
effectively on-line as long as the computations necessary to obtain $\hat s_n$
can be carried out fast enough. 

\section{RESULTS}
\hspace*{\parindent}{\boldmath $NMR$ $laser$ $data$}{\bf---}
As a first example, we show in Fig.~\ref{fig:raser} the result of applying the
described procedure to a data set from an NMR laser experiment~\cite{raser}.
The same data has also been used in Ref.~\cite{buzug}. The laser is
periodically driven and once per driving cycle the envelope of the laser output
is recorded. The resulting sampling rate is 91~Hz. At this rate, the nonlinear
noise reduction scheme can be easily carried out in real time on a Pentium~II
processor at 200~MHz. Since further iterations can be carried out after the
time corresponding to one embedding window without interfering with the
previous steps, we could perform up to three iterations on a dual Pentium~II
workstation at 300~MHz in real time. The figure shows the result after two
iterations. Projections from $m=7$ down to $Q=2$ dimensions were used, at least
100 neighbours were requested at a neighbourhood size of 2~units. The history
was limited to 20000 samples, or 220~s. This fairly large data base is needed
since the initial noise level is already small (less than 2\%~\cite{buzug}) and
small neighbourhoods are required to avoid the dominance of curvature artefacts.

{\boldmath $Magneto$\mbox{\bf -}$cardiogram$ $data$}{\bf---} 
The actual acceleration resulting from the above modifications strongly
depends on the situation and it is difficult to give general rules and
benchmarks. Let us however study a realistic example in some detail to
illustrate the main points. In electrophysiological research, it is quite
attractive to augment the measurements of electric potentials with recordings
of the magnetic field strength. The latter penetrate intervening tissue much
more efficiently. Of particular interest are {\em magneto-encephalographic}
(MEG) recordings which allow to access regions of the brain noninvasively
which cannot be monitored electrically using surface electrodes. In cardiology,
the {\em magneto-cardiogram} (MCG) provides additional information to the
traditional {\em electrocardiogram} (ECG). A particular application is the
noninvasive monitoring of the fetal heart which is otherwise complicated by
shielding of the electric field by intervening tissue. A common problem with
magnetic recordings, however, is that the fields are rather feeble and the
measurements have to be carried out in  a shielded room. Even then, noise
remains a major challenge for this experimental technique. We will demonstrate
in the following how nonlinear noise reduction could be used for continuous 
MCG monitoring. 

\begin{figure}
\centerline{\hspace*{1.5cm}\input{f_mcgd.tex}\hspace*{-3.5cm}\input{f_mcgd_c.tex}}
   \caption[]{Result of nonlinear filtering of a MCG time series}
\end{figure}
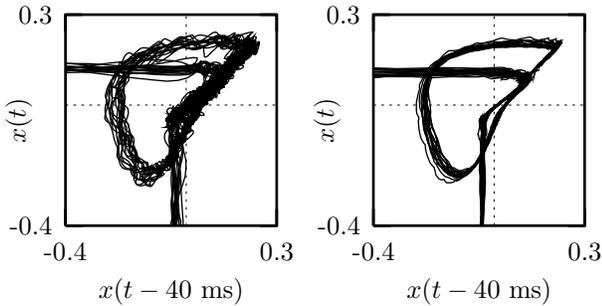

\begin{figure}
\centerline{\hspace*{2cm}\input{f_mcg.tex}\hspace*{-3cm}\input{f_mcg_c.tex}}
   \caption[]{Result of nonlinear filtering of an MCG time series}
\end{figure}
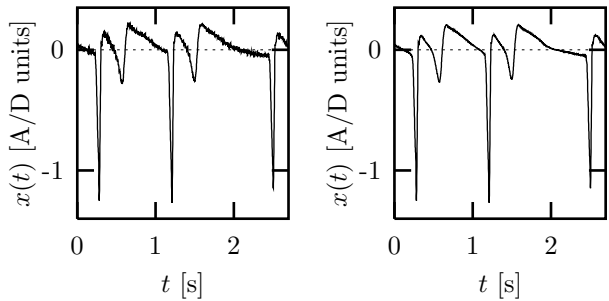

We use an MCG recording of a normal human subject at rest. The data was kindly
provided by Carsten Sternickel at the University of Bonn. The sampling rate was
1000~Hz, which is quite high for cardiac monitoring. For the signal processing
task, any sampling rate above about 200~Hz would be sufficient. (Below 200~Hz,
the spike representing the depolarisation of the ventricle might not be
resolved properly). In order to cover a significant fraction of one cardiac
cycle by an embedding window, we chose an embedding with delay $\tau =10$~ms in
$m=10$ dimensions. Neighbourhoods were formed with a radius of $0.06$
uncalibrated A/D units of the recording, about three times the
estimated noise level. We quote computation times for the processing of 10~s of
MCG on a Pentium processor at 133~MHz, determined on a laptop PC running the
Linux operating system. The timing results are summarised in
Table~\ref{table}. The data base of representatives for (f) was formed by
assuming that the local linear subspaces are equivalent on length scales of the
order of 0.06 A/D units.

\begin{table}
\centerline{
\begin{tabular}{llr}
\hline
a) & all neighbours                   & 490~s \\
b) & box assisted neighbour search    &  94~s \\ 
c) & all neighbours in past           & 246~s \\
d) & $n-k<5$~s                        & 191~s \\
e) & (d) and $U_{\rm max}<200$       &  44~s \\
f) & (e) and reuse of representatives &   6~s \\
\hline
\end{tabular}}
\caption[]{Computation time for nonlinear noise reduction of 10~s of an MCG
  recording sampled at 1000~Hz. See text for details.\label{table}}
\end{table}

{\boldmath $Fetal$ $ECG$ $extraction$}{\bf---} Let us finally discuss the
extraction of the fetal electrocardiogram (FECG) from non-invasive maternal
recordings. Other very similar applications include the removal of ECG
artefacts from electro-myogram (EMG) recordings (electric potentials of muscle)
and spike detection in electro-encephalogram (EEG) data.

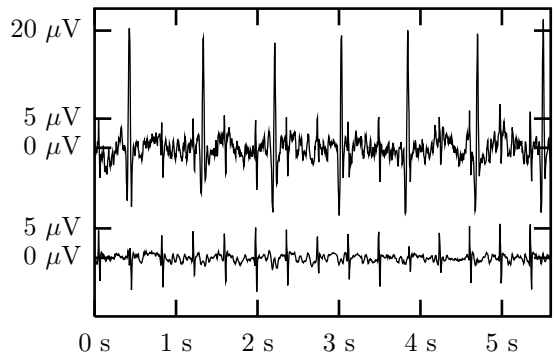
\begin{figure}
   \centerline{\hspace*{-1.5cm}\input{f_fecg.tex}}
   \caption[]{Signal separation by locally linear projections in phase space.
   The original recording (upper trace) contains the fetal ECG hidden under
   noise and the large maternal signal. Projection onto the manifold formed
   by the maternal ECG (middle) yields fetus plus noise, another projection
   yields a fairly clean fetal ECG (lower trace). The data was kindly provided
   by J.~F. Hofmeister~\cite{recording}.\label{fig:fecg}}
\end{figure}

Fetal ECG extraction can be regarded as a three-way filtering problem since we
have to assume that a maternal abdominal ECG recording consists of three main
components, the maternal ECG, the fetal ECG, and exogenous noise, mostly from
action potentials of intervening muscle tissue. All three components have
quite similar broad band power spectra and cannot be filtered apart by
spectral methods. The fetal component is detectable from as early as the
eleventh week of pregnancy. After about the twentieth week, the signal becomes
weaker since the electric potential of the fetal heart is shielded by the {\em
  vernix caseosa} forming on the skin of the fetus. It appears again towards
delivery. In Refs.~\cite{fetal1,fetal2}, it has been proposed to use a
nonlinear phase space projection technique for the separation of the fetal
signal from maternal and noise artefacts. A typical example of output of this
procedure is shown in Fig.~\ref{fig:fecg}. The assumption made about the
nature of the data is that the maternal signal is well approximated by a
low-dimensional manifold in delay reconstruction space. After projection onto
this manifold, the maternal signal is separated from the noisy fetal
component. Now it is assumed that the fetal ECG is also approximated by a
low-dimensional manifold and the noise is removed by projection. Since both
manifolds are curved, the projections have to be made onto linear
approximations. For technical details see Refs.~\cite{fetal1,fetal2}.

\section{DISCUSSION}
We have demonstrated that by certain modifications to the
algorithms described in the literature, nonlinear projective noise reduction
can be turned into a signal processing tool that can in many situations run in
real time in a data stream. The class of problems that can be solved by this
approach is much wider than initially assumed since strict determinism of the
signal is not necessary. Signal and noise have to be distinguishable by their 
shape in a reconstructed phase space.

The algorithms used in this paper and in Ref.~\cite{I} are publicly available 
as part of the TISEAN software project~\cite{tisean}. All the examples were
indeed carried out with these implementations.

We thank Leci Flepp, Carsten Sternickel, and John F. Hofmeister for letting us
use their data. This work was supported by the SFB 237 of the Deutsche
Forschungsgemeinschaft.

\end{document}

%% file: f_raser.tex
\setlength{\unitlength}{0.1bp}
\begin{picture}(1980,1296)(0,0)
\special{psfile=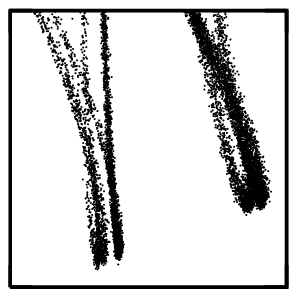 llx=0 lly=0 urx=396 ury=302 rwi=3960}
\put(848,150){\makebox(0,0){$x(t-\mbox{11 ms})$}}
\put(300,798){%
\special{ps: gsave currentpoint currentpoint translate
270 rotate neg exch neg exch translate}%
\makebox(0,0)[b]{\shortstack{$x(t)$}}%
\special{ps: currentpoint grestore moveto}%
}
\put(1246,300){\makebox(0,0){30}}
\put(450,300){\makebox(0,0){15}}
\put(400,1196){\makebox(0,0)[r]{-25}}
\put(400,400){\makebox(0,0)[r]{-40}}
\end{picture}

%% file: f_raser_c.tex
\setlength{\unitlength}{0.1bp}
\begin{picture}(1980,1296)(0,0)
\special{psfile=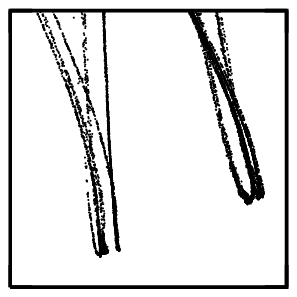 llx=0 lly=0 urx=396 ury=302 rwi=3960}
\put(848,150){\makebox(0,0){$x(t-\mbox{11 ms})$}}
\put(300,798){%
\special{ps: gsave currentpoint currentpoint translate
270 rotate neg exch neg exch translate}%
\makebox(0,0)[b]{\shortstack{$x(t)$}}%
\special{ps: currentpoint grestore moveto}%
}
\put(1246,300){\makebox(0,0){30}}
\put(450,300){\makebox(0,0){15}}
\put(400,1196){\makebox(0,0)[r]{-25}}
\put(400,400){\makebox(0,0)[r]{-40}}
\end{picture}

%% file: f_mcgd.tex
\setlength{\unitlength}{0.1bp}
\begin{picture}(2087,1296)(0,0)
\special{psfile=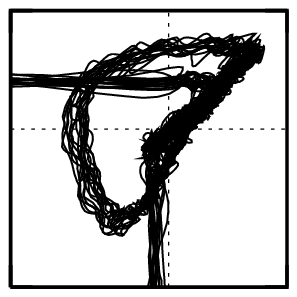 llx=0 lly=0 urx=418 ury=302 rwi=4180}
\put(898,150){\makebox(0,0){$x(t-\mbox{40 ms})$}}
\put(350,798){%
\special{ps: gsave currentpoint currentpoint translate
270 rotate neg exch neg exch translate}%
\makebox(0,0)[b]{\shortstack{$x(t)$}}%
\special{ps: currentpoint grestore moveto}%
}
\put(1296,300){\makebox(0,0){0.3}}
\put(500,300){\makebox(0,0){-0.4}}
\put(450,1196){\makebox(0,0)[r]{0.3}}
\put(450,400){\makebox(0,0)[r]{-0.4}}
\end{picture}

%% file: f_mcgd_c.tex
\setlength{\unitlength}{0.1bp}
\begin{picture}(2087,1296)(0,0)
\special{psfile=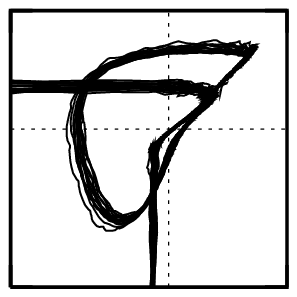 llx=0 lly=0 urx=418 ury=302 rwi=4180}
\put(898,150){\makebox(0,0){$x(t-\mbox{40 ms})$}}
\put(350,798){%
\special{ps: gsave currentpoint currentpoint translate
270 rotate neg exch neg exch translate}%
\makebox(0,0)[b]{\shortstack{$x(t)$}}%
\special{ps: currentpoint grestore moveto}%
}
\put(1296,300){\makebox(0,0){0.3}}
\put(500,300){\makebox(0,0){-0.4}}
\put(450,1196){\makebox(0,0)[r]{0.3}}
\put(450,400){\makebox(0,0)[r]{-0.4}}
\end{picture}

%% file: f_mcg.tex
\setlength{\unitlength}{0.1bp}
\begin{picture}(1980,1296)(0,0)
\special{psfile=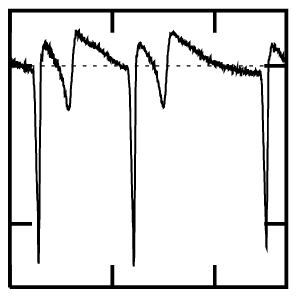 llx=0 lly=0 urx=396 ury=302 rwi=3960}
\put(798,150){\makebox(0,0){$t$ [s]}}
\put(219,798){%
\special{ps: gsave currentpoint currentpoint translate
270 rotate neg exch neg exch translate}%
\makebox(0,0)[b]{\shortstack{$x(t)$ [A/D units]}}%
\special{ps: currentpoint grestore moveto}%
}
\put(990,300){\makebox(0,0){2}}
\put(695,300){\makebox(0,0){1}}
\put(400,300){\makebox(0,0){0}}
\put(350,1037){\makebox(0,0)[r]{0}}
\put(350,582){\makebox(0,0)[r]{-1}}
\end{picture}

%% file: f_mcg_c.tex
\setlength{\unitlength}{0.1bp}
\begin{picture}(1980,1296)(0,0)
\special{psfile=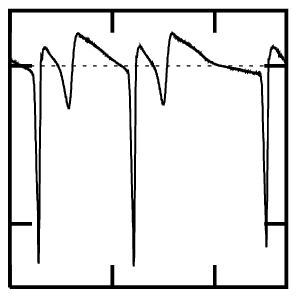 llx=0 lly=0 urx=396 ury=302 rwi=3960}
\put(798,150){\makebox(0,0){$t$ [s]}}
\put(219,798){%
\special{ps: gsave currentpoint currentpoint translate
270 rotate neg exch neg exch translate}%
\makebox(0,0)[b]{\shortstack{$x(t)$ [A/D units]}}%
\special{ps: currentpoint grestore moveto}%
}
\put(990,300){\makebox(0,0){2}}
\put(695,300){\makebox(0,0){1}}
\put(400,300){\makebox(0,0){0}}
\put(350,1037){\makebox(0,0)[r]{0}}
\put(350,582){\makebox(0,0)[r]{-1}}
\end{picture}

%% file: f_fecg.tex
\setlength{\unitlength}{0.1bp}
\begin{picture}(2519,1511)(0,0)
\special{psfile=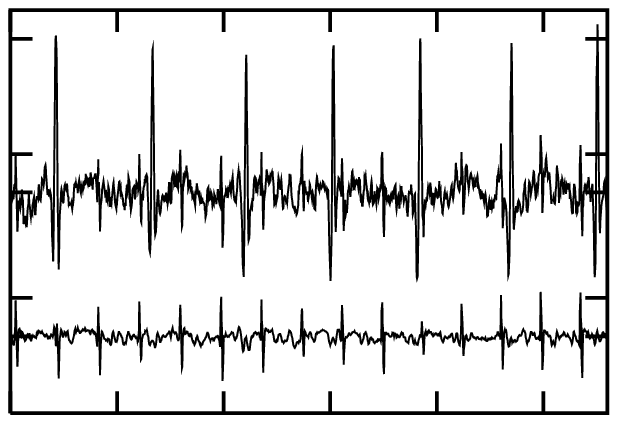 llx=0 lly=0 urx=504 ury=353 rwi=5040}
\put(2285,150){\makebox(0,0){5 s}}
\put(1978,150){\makebox(0,0){4 s}}
\put(1671,150){\makebox(0,0){3 s}}
\put(1364,150){\makebox(0,0){2 s}}
\put(1057,150){\makebox(0,0){1 s}}
\put(750,150){\makebox(0,0){0 s}}
\put(700,582){\makebox(0,0)[r]{5 $ \mu $V}}
\put(700,471){\makebox(0,0)[r]{0 $ \mu $V}}
\put(700,1328){\makebox(0,0)[r]{20 $ \mu $V}}
\put(700,996){\makebox(0,0)[r]{5 $ \mu $V}}
\put(700,886){\makebox(0,0)[r]{0 $ \mu $V}}
\end{picture}